# Fiber-Optic-Gyroscope Measurements Close to Rotating Liquid Helium

M. Tajmar and F. Plesescu

*Space Propulsion & Advanced Concepts*
*Austrian Institute of Technology GmbH*
*A-2444 Seibersdorf, Austria*
*+43-50550-3142; martin.tajmar@ait.ac.at*

**Abstract.** We previously reported anomalous fiber-optic gyroscope signals observed above rotating rings at temperatures close to liquid helium. Our results suggested that the liquid helium itself may be the source of our observed phenomenon. We constructed a new cryostat experiment that allows rotating a large quantity of liquid helium together with a superconducting niobium tube. The facility is built in such a way that our gyroscope can be placed directly in the center of rotation along the axis; however, the cryostat is built around the gyroscope to allow measuring without interference of helium liquid or gas. An anomalous signal was found of similar value compared to our previous measurements with a changed sign. As this measurement was done at a different location (center position) with respect to our old setup (top position), first hints for a possible field distribution of this phenomenon can be made. However, due to lower angular velocities used in this new setup so far, our measurement resolution was close to three times the resolution of our gyroscope and hence our data represent work in progress.

**Keywords:** Frame Dragging, Liquid Helium, Superfluidity, Superconductivity

**PACS:** 04.80.-y, 74.00.00

## INTRODUCTION

Gravity is the weakest of all four fundamental forces. Since Einstein's general relativity theory from 1915, we know that gravity is not only responsible for the attraction between masses but that it is also linked to a number of other effects such as bending of light or slowing down of clocks in the vicinity of large masses. One particularly interesting aspect of gravity is the so-called Thirring-Lense or Frame-Dragging effect: A rotating mass should drag space-time around it, affecting for example the orbit of satellites around the Earth. However, the effect is so small that it required the analysis of 11 years of satellite orbit data to confirm Einstein's prediction within ±10% (Ciufolini and Pavlis, 2004).

In principle, such frame-dragging fields could be used to generate artificial gravitational fields, as already proposed by Forward (1963), however it requires the mass densities of a neutron star to generate gravitational effects on the scale of the Earth's gravitational field. Therefore, apart from Newton's mass attraction, gravity is believed to be only accessible via astronomy or satellite experiments but not in a laboratory environment.

That assumption was recently challenged (Tajmar and de Matos, 2003; 2005), proposing that a large frame-dragging field could be responsible for a reported mass anomaly found in niobium superconductors (Tate *et al*, 1989; 1990). Using grants from the US Air Force Office for Scientific Research (AFRL/AFOSR), the European Space Agency (ESA) and the Austrian Institute of Technology (AIT), several experiments were setup during the past 6 years to investigate if such large frame-dragging fields indeed exist in the vicinity of rotating matter at low temperatures using laser gyroscopes and very sensitive accelerometers (Tajmar *et al*, 2006; 2007; 2008; 2009a).

Our last experiment consisted of a ring or small container made out of various materials, which is cooled by liquid helium (Tajmar *et al*, 2009a). Above the ring, a military-grade fiber-optic-gyroscope is mounted on a separate structure that remains at rest. While rotating the ring with a cryomotor and after passing a critical temperature measured inside the ring, the gyro senses rotation. A number of different setups led to the following observations:

1. The effect is independent of the ring's material including normal metals, superconductors or plastic.
2. The gyro's signal appears to be proportional to the ring's angular velocity with a coupling factor in the range of $10^{-8}$.
3. All setups showed a parity violation, i.e. clockwise and counter-clockwise rotation yielded different results. In some setups, only clockwise rotation showed the effect (Setup A and C) and another setup (Setup B) showed the same signal in the same direction regardless of the sense of rotation.
4. In Setups A and B, the gyroscope was mounted inside a separately mounted vacuum chamber in close connection to the low temperature rotating ring. In Setup C, the rotating ring assembly was totally isolated inside a cryostat and the gyroscope was mounted outside the facility. Here, the signal strength decreased by more than an order of magnitude.

This leads to the conclusion that the origin of the effect is either connected to a critical low temperature independent of the material used or that the effect is trigged by the cryogenic liquid helium itself. Some small part of our relatively large liquid helium volume may have been in the superfluid state due to thermal fluctuations. Chiao (1982) and Tajmar and de Matos (2003; 2005) proposed that rotating superfluid helium may be the source of large frame-dragging-like fields in order to explain its resistance to rotation similar to the magnetic fields developed by rotating superconductors that keep the Cooper-pairs at rest. Initial experiments by Moulthrop (1984) exclude the inertial coupling of two separate superfluids down to a coupling factor of 0.05 which is, however, some 7 orders of magnitude above our observed anomaly.

The parity violation on the other hand points to either a systematic effect that was not recognized so far or that the origin of the effect is outside of the known four fundamental forces of nature. The signal reduction in Setup C may also point towards systematic effects such as acoustic vibration in Setups A and B, however, also in the case of much improved isolation as in Setup C, an anomalous effect on the gyro remains although close to our measurement resolution.

In order to further investigate the real origin behind our signals, we decided to design a new experimental setup which allows rotating liquid and superfluid helium in a controlled manner and to continue mapping the anomalous gyro effect outside of the facility at different locations to study its expansion.

## EXPERIMENTAL SETUP

The main part of our experiment is a rotating pot as shown in Figure 1 inside a large custom-built cryostat which can be evacuated as shown in Figure 2. The cryostat is mounted on a structure which allows tilting it along the Earth's north-south axis (we plan to tilt the cryostat in future experiments to investigate the influence of the Earth's spin). All experiments performed so far were done with the cryostat mounted vertically to our laboratory floor. The pot inside can be filled with up to 30 L of liquid helium. Its outer and inner radii are 211 mm and 93.5 mm and the outer and inner heights are 271 mm and 177 mm respectively. The pot can only be filled through a ring shaped hole on the top and it has four fins that ensure that the whole liquid is homogenously rotating with the same velocity. All outer pot surfaces are made out of a stainless steel with a thickness of 0.5 mm, only the fins inside and the axis to the motor are made out of aluminum. In addition, a niobium cylinder with the same thickness was inserted and glued on the inner pot cylinder in order to include a rotating superconductor in our tests as well as to provide an additional magnetic shield for the location of the gyroscope.

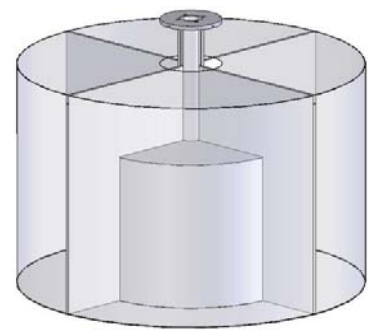

**FIGURE 1.** Rotating Pot Filled with 30 L of Liquid Helium (Outer Wall is Transparent)

The pot is mechanically attached to the axis of a cryomotor (Phytron VSH-UHVC 100) allowing it to spin up to an angular velocity of 60 rad/s. Higher speeds were prohibited due to the weight of the pot and the liquid helium on the motor low temperature bearings. Silicon diodes (Lakeshore DT-670B-SD) were placed in the middle of the inner cylindrical surface of the pot (readout via a collector ring as it is rotating with the pot) and on the top close to the cryomotor in order to monitor the filling level of the liquid helium and the temperature.

We used a military-grade OPTOLINK SRS-1000 fiber-optic-gyroscope, which is mounted on a support structure outside the cryostat allowing it to change its position and orientation all around the cryostat. The cryostat has such a shape, that the gyroscope can be inserted along the middle-axis up to the center of the rotating pot (see Figure 2). In addition, a magnetic field sensor (Honeywell SS495A1), high resolution accelerometers (Colibrys SiFlex1500 and SiliconDesigns 1221) and temperature sensors are mounted together with the gyroscope. All sensors are encapsulated under a high permeability magnetic shield (permalloy) in order to reduce their magnetic sensitivity.

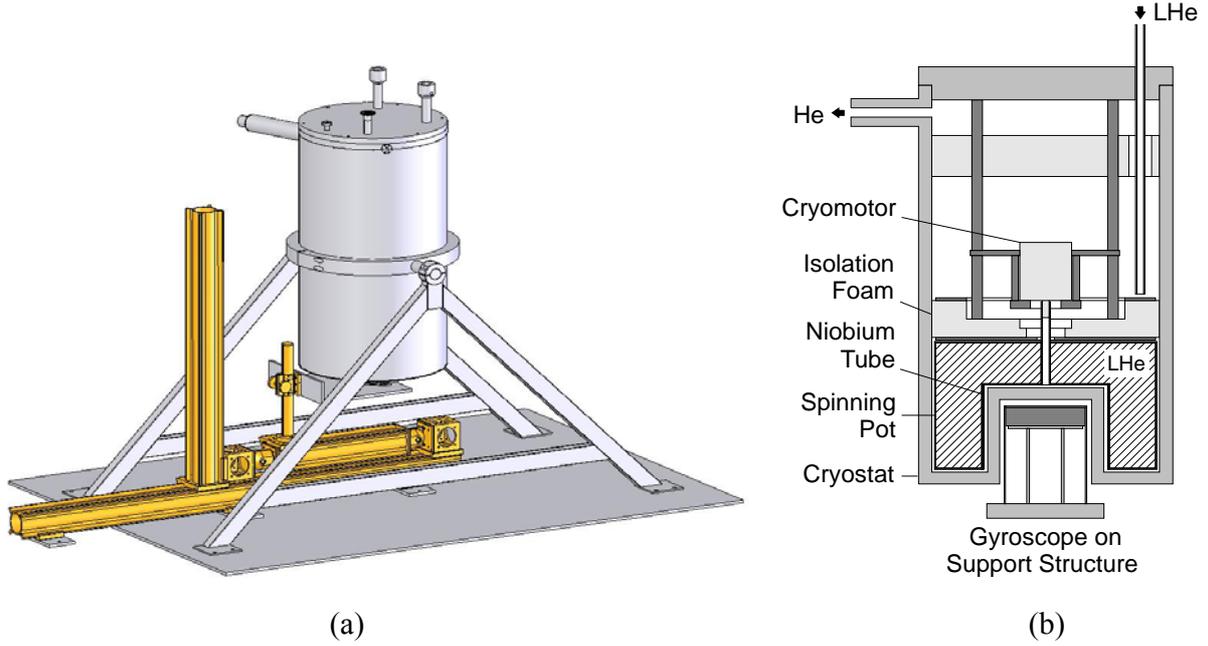

**FIGURE 2.** Experimental Setup D: (a) Overall Assembly of Cryostat and Gyroscope Support Structure, (b) Detailed Intersection.

## Sensitivity and Systematic Effects

Since we are trying to measure very small rotation rates, it is important to know the minimum rotation rate that can be resolved by the fiber-optic gyroscope. For the SRS-1000 gyroscope, the upper limit of the minimal measured rotation rate, caused by polarization nonreciprocity of a light source with a Gauss spectrum, is expressed as (Prilutskii, 2003; OPTOLINK - personal communication with V. A. Fedorov)

$$\Omega_{min} \leq \frac{\lambda c p \varepsilon}{DL} \sqrt{\frac{\lambda \sqrt{\ln 2} h L_p}{\pi \Delta \lambda}} \quad (1)$$

where $\lambda=1.55\times10^{-6}$ m is the average wavelength of light, $\Delta\lambda=50\times10^{-9}$ m is the width of the light spectrum, $c$ is the speed of light in vacuum, $D=150\times10^{-3}$ m is the diameter of the fiber coil, $L=1070$ m is the length of the fiber coil, $h=1\times10^{-6}$ m$^{-1}$ is the polarization crosstalk of the fiber coil, $L_p=2.5\times10^{-3}$ m is the beat length of the fiber coil, $p$ is the residual degree of polarization of the light source and $\varepsilon$ is the coefficient of the polarizer's extinction. For typical values of $p\approx0.01$ and $\varepsilon \approx0.01$ (i.e. 40 db), equation (1) gives an upper limit of $\Omega_{min}\leq4.1\times10^{-8}$ rad/s. This coincides with the bias drift value specified for the SRS-1000. We also tested the gyroscope on a piezo-activated nano-rotation table (Tajmar et al, 2009b) and could resolve velocity steps of $1\times10^{-7}$ rad/s with an accuracy of $1.5\times10^{-8}$ rad/s which is just below the theoretically predicted minimum rotation rate upper limit. No asymmetry has been seen between clockwise or counter-clockwise gyro responses.

The magnetic sensitivity along the gyro's axis is $4.8\times10^{-3}$ rad/s/T. According to our Hall sensor mounted directly on the gyroscope, the magnetic field changes during rotation (influence from stepper motor and London magnetic field from rotating superconducting niobium cylinder) were always less than 1 µT. Therefore, the magnetic influence is about one order of magnitude below our minimum rotation rate.

Vibration is always a source for systematic effects although it is stated frequently by manufacturers that fiber-optic-gyroscopes are insensitive to vibration. For the SRS-1000 gyro, a vibration environment of 20-2000 Hz with 2 g rms is specified for normal operation. These are severe vibrations and correspond to the environment faced for equipment during the launch of satellites. Our gyroscope is mounted on a structure fixed to the laboratory floor and is therefore only exposed to normal seismic noise. The stepper motor is also producing acoustic noise in higher frequency bands, however, here it makes no difference if the motor is working at low or normal temperatures and therefore such influence can be easily assessed in room temperature calibration measurements. We have seen no influence by the stepper motor and the rotating pot at our calibration runs so far.

One of the weakest points of our setup was that we only commanded the stepper motor velocity without feedback of the actual velocity and spin orientation. We recently implemented a magnetic encoder that we will use for future measurements.

## EXPERIMENTAL RESULTS

The following data was obtained following a standardized speed profile which lasts about 210 seconds. The profile is subdivided into 5 sectors with pre-defined time intervals: rest, acceleration, maximum speed, de-acceleration and rest. Each profile is performed in clockwise direction (spin vector points downwards) and in counter-clockwise direction. Random noise was reduced by applying a 200 pt digital moving average (DMA) filter to the gyro output and motor velocity. A sampling rate of 13 Hz ensured that the averaging window is always smaller than the maximum speed time interval of the profile. The noise was further reduced by signal averaging of at least 20 profiles for each test case which also increases the statistical significance of our results. All results are summarized in Table 1.

**Table 1.** Summary of Experimental Data.

| Rotating Fluid | Gyroscope Position | Coupling Factor (Gyro Output/$\omega$) | |
|---|---|---|---|
| | | CW | CCW |
| Liquid Nitrogen | Center | $-0.3\pm0.2\times10^{-9}$ | $0.6\pm0.2\times10^{-9}$ |
| Liquid Helium | Center | $-1.7\pm0.4\times10^{-9}$ | $-0.2\pm0.3\times10^{-9}$ |
| | Bottom | $-0.1\pm0.2\times10^{-9}$ | $-0.3\pm0.3\times10^{-9}$ |
| | Side | $1.0\pm0.3\times10^{-9}$ | $-0.5\pm0.3\times10^{-9}$ |
| Superfluid Helium* $^4$He | Center | $1.9\pm3.4\times10^{-9}$ | $0.1\pm2.0\times10^{-9}$ |

*Non-quantized flow due to fins inside rotating pot

## Liquid Helium Measurements

Initially, liquid helium is filled into the cryostat until it reaches the cryomotor to be sure that the whole rotating pot is filled with liquid helium. We measured the gyro response at three different locations as outlined in Figure 3a in order to test if the anomalous signal behaves like a dipole outside of the cryostat as theoretically expected from frame-dragging fields. Figure 3b shows the gyro measurement at the center position (1) when the pot was rotating with angular velocities up to 60 rad/s. For comparison, plots are shown both when the pot was filled with liquid helium and liquid nitrogen which was our reference case. Figure 4 shows the gyro measurements for rotating liquid helium at position (2) and (3). The noise level is close to the gyro's minimum rotation rate upper limit of $\pm4\times10^{-8}$ rad/s. We can summarize our observations as follows:

1. While rotating of liquid nitrogen does not show any anomalous behavior beyond the gyro's minimum rotation rate, liquid helium does appear to affect the gyroscope.

2. At the center position (1), the coupling factor between gyro response and applied angular velocity ω to the rotating pot is (Gyro Output/ω)=-1.7±0.4×10$^{-9}$ which is nearly identical to the measurement in our old facility with Setup C (Tajmar *et al*, 2009a) which was (Gyro Output/ω)=1.7±0.3×10$^{-9}$ but with the major difference that the sign has flipped as one would expect from a dipolar field distribution. This is a very important point as we have always performed measurements with the gyro above the rotating ring/liquid so far. Note that there is again a parity anomaly such that the effect is pronounced in the clockwise-direction as previously observed in Setups A and C.

3. Position (2) did not show an anomalous effect within our sensitivity and at position (3) we observed an effect about 40% reduced from position (1) and with a change in sign. All this would be expected from a dipolar field distribution of our anomalous effect as illustrated in Figure 3a. However, this signal is already very close to our minimum rotation rate upper limit and may well be noise.

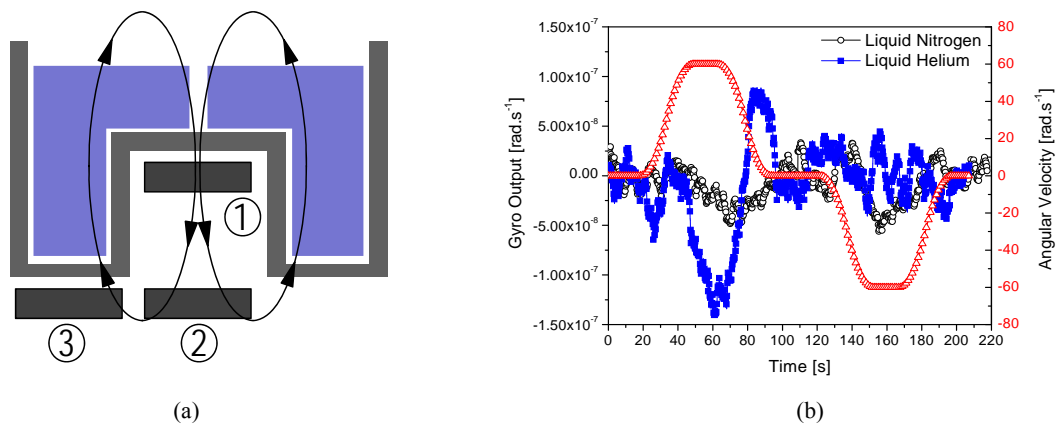

(a)          (b)

**FIGURE 3.** (a) Position of the Gyroscope (1… Center, 2 … Bottom and 3 … Side), (b) SRS Gyro Output for Liquid Helium and Liquid Nitrogen Versus Applied Angular Velocity (Δ) at Position 1.

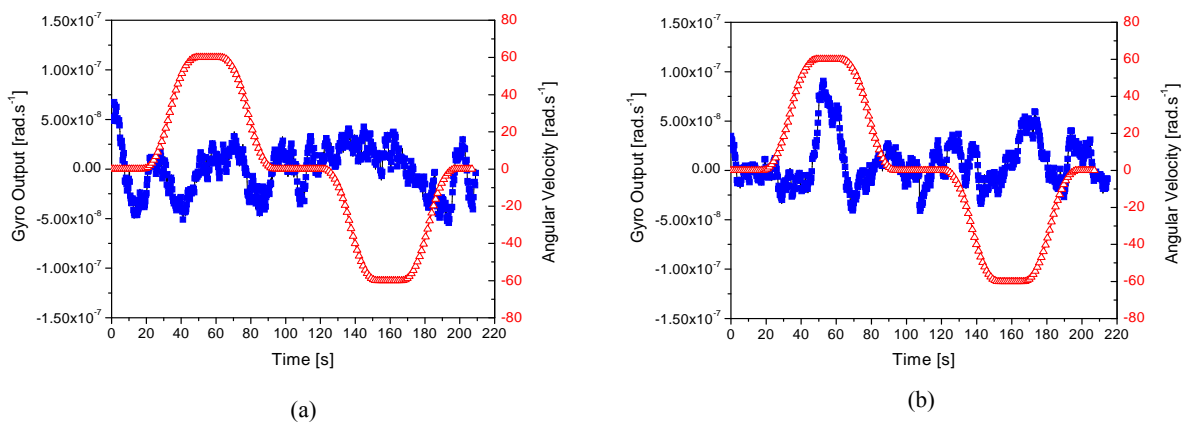

(a)          (b)

**FIGURE 4.** SRS Gyro Output for Liquid Helium (a) at Position 2 and (b) Position 3.

Unfortunately, the anomalous signal at the center position (1) is only about three times above our minimum rotation rate upper limit. Higher angular velocities are clearly necessary in order to obtain better results and to clarify the origin of our gyroscope responses. As the inner wall of the rotating pot was covered with a niobium cylinder, our results may be also used as a limit for anomalous frame-dragging fields from rotating superconductors where the gyroscope was placed in the center of the rotating cylinder.

# Preliminary Superfluid Helium Measurements

In order to further study the origin of the effect, we decided to pump on the liquid helium in order to create superfluid helium $^4$He. Since the volume reduces during pumping (about 35% down to 1.5 K), we had to fill in about 1.5 times the amount of liquid helium (45 L) compared to our previous effort so that the liquid helium level was initially above the cryomotor. This large amount of liquid helium required a sufficiently large vacuum pump due to the large boil-off from the liquid. We had to use a Leybold Sogevac SV300 with a pumping speed of 300 m³/h in order to achieve superfluid temperatures below the Lambda point of 2.17 K. After 30 minutes, the silicon diodes in the center and at the top of the rotating pot indicated 1.3 and 1.9 K respectively. We only performed preliminary measurements so far with a maximum angular velocity of 10 rad/s with the gyroscope at the center position (1) as shown in Figure 5.

No signal beyond the noise level can be identified. However, the angular velocity was reduced by a factor of 6 compared to our previous measurements. It's very important to remember that due to the fins inside our rotating pot, no quantization of angular momentum was possible which is predicted by many theories to be the underlining cause of large frame-dragging fields. Future experiments will remove part of the fins in order to allow the superfluid to close the loop around the pot and to form vortices. We are also planning to increase the angular velocity in order to improve the sensitivity of the coupling factor between the gyro and the rotating superfluid.

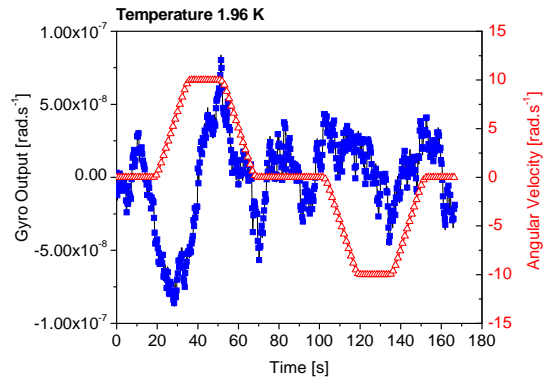

**FIGURE 5.**   SRS Gyro Output for Superfluid Helium at Center Position (1).

# CONCLUSION

Fiber-optic-gyroscope signals were measured outside of a specially-shaped cryostat which contains a rotating pot that was filled with liquid nitrogen, liquid helium and superfluid helium. For the first time we could measure at different locations such as at the center, bottom and side of the rotating pot. Rotating liquid helium produced similar anomalous results as observed in previous measurements with a different setup (Setup C) but for the first time with a change in sign at the center position with respect to our old position on top which is expected from a dipolar field distribution. The sign changed again when the gyro was positioned below the rotating pot. First results were reported for rotating superfluid helium, however, we only used very small angular velocities and our rotating pot had fins inside which prohibit the creation of vortices and quantized flow.

Our measurements outside the cryostat so far are still work in progress and the anomalous signals are just three times above or even below the minimum rotation rate upper limit of the gyroscope used and hence remain doubtful. We plan to implement a magnetic encoder in future measurements to validate the speeds of the rotating pot as well as to go to higher angular velocities with a new cryomotor to significantly increase our sensitivity.

# ACKNOWLEDGMENTS

This work was funded by the Austrian Institute of Technology GmbH. We gratefully acknowledge our continuing and inspiring discussions with Clovis J. de Matos.